\documentclass[twocolumn,trackchanges]{aastex631} %twocolumn,linenumbers,

\usepackage{amsmath}
\usepackage{multirow}
\usepackage{makecell}
\usepackage{tabularx}
\usepackage{xcolor}
\usepackage{mathrsfs}
\usepackage{CJKutf8} 
\usepackage{graphicx}
\usepackage{hyperref} 
\usepackage{enumitem}
\usepackage{booktabs}
\usepackage{anyfontsize}
\usepackage{bm}
\usepackage{amsmath}

\begin{document}
\begin{CJK*}{UTF8}{gbsn}
    \title{Search for neutrino emission from LHAASO observed Microquasar with IceCube 10-year data}
    
    \correspondingauthor{Hao-Ning He}
    \email{hnhe@pmo.ac.cn}
    
    \author[0009-0002-9585-0210]{Rong-Lan Li (李荣岚)}
    \affiliation{Key Laboratory of Dark Matter and Space Astronomy, \\
    Purple Mountain Observatory, Chinese Academy of Sciences, Nanjing 210023, China}
    \affiliation{School of Astronomy and Space Sciences, University of Science and Technology of China, Hefei, 230026, Peopleʼs Republic of China}
    
    \author[0000-0002-8941-9603]{Hao-Ning He* (贺昊宁)}
    \affiliation{Key Laboratory of Dark Matter and Space Astronomy, \\
    Purple Mountain Observatory, Chinese Academy of Sciences, Nanjing 210023, China}
    \affiliation{School of Astronomy and Space Sciences, University of Science and Technology of China, Hefei, 230026, Peopleʼs Republic of China}
    
    \author[0000-0002-9758-5476]{Da-Ming Wei (韦大明)}
    \affiliation{Key Laboratory of Dark Matter and Space Astronomy, \\
    Purple Mountain Observatory, Chinese Academy of Sciences, Nanjing 210023, China}
    \affiliation{School of Astronomy and Space Sciences, University of Science and Technology of China, Hefei, 230026, Peopleʼs Republic of China}
    
    \begin{abstract}
        %Microquasars are potential Galactic cosmic-ray accelerators. 
        %The accelerated cosmic rays are expected to interact hadronically with nearby ambient gas or the interstellar medium, producing high-energy $\gamma$-rays and neutrinos. 
        %The Large High Altitude Air Shower Observatory (LHAASO) has detected $>$100 TeV $\gamma\text{-}$ray emission from five microquasars, indicating their capability to accelerate cosmic rays up to PeV energies. At these energies, Klein-Nishina suppression significantly reduces the efficiency of leptonic emission mechanisms, and neutrino detection would serve as unambiguous evidence of hadronic acceleration. Here, we present the result of a search for neutrino emission from these sources, including a stacking analysis of all known microquasars. No statistically significant signal is found. Based on the derived upper limits, we constrain the hadronic contribution to the observed $\gamma$-ray flux from these sources. We also provide prospects for future observations of these microquasars with upcoming neutrino telescopes.
        The Large High Altitude Air Shower Observatory (LHAASO) has detected ultra-high-energy (UHE; $E > 100$ TeV) gamma-ray emission from five microquasars, suggesting their potential as Galactic PeV cosmic-ray accelerators. At these energies, the Klein–Nishima effect strongly suppresses leptonic processes, making neutrino observation a crucial test for hadronic acceleration. We present a search for neutrino emission from these LHAASO-identified microquasars using ten years of IceCube muon-track data. No significant neutrino signal was found in either single-source or stacking analyses. %From the derived upper limits, we constrain the hadronic contribution to the observed gamma-ray flux for each source, considering both simple power-law models and dedicated physical scenarios. 
        Our stacking result further shows that the studied microquasar population can only account for a small fraction of the diffuse neutrino flux along the Galactic Plane. Finally, we demonstrate that next-generation neutrino telescopes, such as HUNT, will have the sensitivity to probe hadronic emission from these candidate PeVatrons.
    \end{abstract}
    
    
    \section{Introduction} \label{sec:intro}
    %Since the discovery of cosmic rays (CRs) by Victor Hess in 1912 \citep{hess1912}, their origin and acceleration have remained a persistent puzzle in astrophysics \citep{Kotera_2011,Kachelrie__2019}. The cosmic ray spectrum follows a power law of $E^{-2.7}$ up to the knee at $E_{\rm CR} \sim 3\,\mathrm{PeV}$, above which it steepens \citep{TIBETIII:2008qon,LHAASO:2024knt}. CRs below the knee are widely considered to be of Galactic origin \citep{Hillas:2005cs}, implying the presence of PeVatrons in the Milky Way capable of accelerating particles to $1\,\mathrm{PeV}$. 
    %Deflections by galactic magnetic fields prevent cosmic-ray arrival directions from pointing back to their sources, hindering direct identification of their origin. 
    %Fortunately, cosmic rays interact with the ambient gas and radiation fields, producing secondary particles. Neutral pions ($\pi^0$) decay into gamma rays, while charged pions ($\pi^\pm$) decay into neutrinos. These neutral particles are not deflected by magnetic fields, making them valuable tracers of cosmic-ray acceleration sites \citep{Halzen_2002}. 

The origin of Galactic cosmic rays, especially the PeVatrons responsible for accelerating particles up to the PeV regime (the "knee" in the spectrum), remains a key question in astrophysics. The arrival directions of charged cosmic rays are randomized by their diffusive transport through the Galactic magnetic field, preventing direct identification of their sources. Instead, we rely on neutral secondaries—gamma rays and neutrinos—produced when cosmic rays interact with ambient matter or radiations. Gamma rays from the neutral pion decay and the accompanying neutrinos from the charged pion decay provide unambiguous evidence of hadronic acceleration \citep{Halzen_2002}.

Microquasars, X-ray binary systems hosting a compact object and a companion star, offer a unique window into relativistic jet physics and accretion dynamics. Their relative proximity enables high-resolution, multiwavelength studies of structural components such as jets and accretion flows, making them essential natural laboratories for these phenomena. In these systems, the accreted material forms a turbulent disk \citep{Mirabel:1999fy, Remillard:2006fc}, and the coupled accretion-ejection system acts as a powerful engine capable of accelerating particles, both in the accretion disk and corona, and within the jets and winds launched \citep{Drury:1983zz, Hoshino:2012xm, Middleton:2018xiq, Kuze:2025wda}.

Recently, the LHAASO Collaboration has identified ultra-high-energy (UHE; $E > 100~\mathrm{TeV}$) gamma-ray emission from five microquasars within its field of view, each exhibiting clear black hole–jet structures \citep{LHAASO:2024psv}. These galactic analogues of active galactic nuclei have long been proposed as potential sites of high-energy neutrino production \citep{Mannheim_1992}. Critically, at energies above $\sim 100~\mathrm{TeV}$, the Klein–Nishina effect strongly suppresses inverse Compton scattering, making leptonic emission mechanisms inefficient. The detected UHE gamma rays thus strongly favor a hadronic origin, implying that microquasars can accelerate cosmic rays to PeV energies. The detection of neutrinos from these sources would thus provide smoking-gun evidence for hadronic acceleration.

Building upon the compelling hadronic interpretation of the LHAASO UHE gamma-ray detections, we present a systematic search for neutrino emission from these microquasar sources. Intriguingly, the positional coincidence between one of these sources and the IceCube neutrino alert IC250813A\footnote{\url{https://gcn.gsfc.nasa.gov/notices_amon_g_b/141240_9390028.amon}}  provides an additional, though tentative, observational clue. To robustly test the hypothesis of hadronic particle acceleration, we performed both single-source and stacking analyzes using ten years of IceCube muon-track data. This paper is organized as follows. Section~\ref{sec:data} describes the IceCube and LHAASO datasets and the source sample, Section~\ref{sec:analyse} details the unbinned likelihood method, Section~\ref{sec:results} presents the results and constraints, and Section~\ref{sec:outlook} provides a summary and discusses future prospects.

%Intriguingly, the error region of the IceCube high-energy neutrino alert IC250813A\footnote{\url{https://gcn.gsfc.nasa.gov/notices_amon_g_b/141240_9390028.amon}} spatially overlaps with one of the LHAASO-detected microquasars. This spatial coincidence suggests a possible association and provides immediate impetus to systematically search for neutrino emission from this source class.
    
%A targeted search for neutrino emission from the LHAASO-detected UHE microquasars can therefore serve as a critical test to distinguish between leptonic and hadronic gamma-ray production mechanisms. In this work, we conduct both single-source and stacking analyses to probe the hadronic origin of gamma rays from these candidate Galactic PeVatrons. Section~\ref{sec:data} describes the IceCube Neutrino Observatory, the LHAASO detector arrays, the neutrino data set, and the microquasar list observed by LHAASO. Our analysis methods are described in Section~\ref{sec:analyse}, and the results are presented in Section~\ref{sec:results}. In Section~\ref{sec:outlook}, we summarize our work and discuss the detection prospects of future detectors.
    
    \section{Data Sets from IceCube and LHAASO}\label{sec:data}
    \subsection{IceCube Neutrino Observations}\label{subsec:IceCube-data}
    The IceCube Neutrino Observatory is a cubic-kilometer detector at the geographic South Pole, composed of 5,160 Digital Optical Modules \citep{IceCube:2008qbc}. It detects neutrinos through Cherenkov radiation emitted by relativistic secondary charged particles generated in neutrino interactions \citep{IceCube:2010dpc}. This analysis utilizes the 10-year public muon track data set\footnote{\url{http://doi.org/DOI:10.21234/sxvs-mt83}} \citep{IceCube:2021xar}, which comprises 1,134,450 events recorded between April 2008 and July 2018. %Muon track, resulting from charged-current (CC) $\nu_\mu$ interactions, provide excellent pointing capabilities, with a median angular resolution of $\lesssim 1^\circ$ for $\sim$TeV events. 
    Each event in the data set is characterized by its arrival time $\left ( t \right )$, reconstructed direction $\left( \alpha, \delta \right)$, angular uncertainty $\left( \sigma \right)$, and reconstructed energy $\left( E_{\rm{{rec}}} \right)$. To account for the evolving detector performance during its construction, the data is categorized into five operational periods: IC40, IC59, IC79, IC86-I, and IC86-II-VII, where the number indicates the detector strings deployed. Dedicated instrument response functions are provided for each period, including the effective area $A_{\rm{eff}}(E_\nu,\delta_\nu)$ and the smearing function $\mathcal{M}(E_{\rm{rec}}|E_\nu,\delta_\nu)$. This data set has been extensively used in searches for point-like neutrino sources \citep{IceCube_10yr_integrated,li_ATLAS17jrp} and catalog analyses \citep{ZhouBei_radioAGN,LiRonglan_catalog}. Building on these established methods, we apply this framework to perform a targeted search for neutrino emission from LHAASO-identified microquasars.
    
    \subsection{LHAASO $\gamma$-ray Observations of Microquasars}\label{subsec:MQ-data}
    The Large High Altitude Air Shower Observatory (LHAASO) is a hybrid extensive air shower array designed for $\gamma$-ray and cosmic ray studies. Although it comprises three main components (the Kilometer Square Array (KM2A), the Water Cherenkov Detector Array (WCDA) and the Wide Field-of-view Cherenkov Telescope Array (WFCTA)), this work primarily utilizes observations from KM2A. Covering $1.3~\rm km^2$ with its electromagnetic particle detectors (ED) and muon detectors (MD), KM2A focuses on Galactic $\gamma$-ray sources above 30 TeV in the northern celestial hemisphere. It achieves an exceptional point source sensitivity of $10^{-14}\mathrm{erg/cm^{2}/s^{1}}$ for photon energies around 100 TeV.
    
    Using this high sensitivity, the LHAASO Collaboration recently reported the detection of UHE $\gamma$-rays from five microquasars, SS 433, V4641 Sgr, GRS 1915+105, MAXI J1820+070 and Cygnus X-1. All of these sources exhibited significant $\gamma$-ray excesses with post-trial significance above $4\sigma$ in the $E > 25~\mathrm{TeV}$ band, indicating the presence of highly efficient particle accelerators within these systems. A summary of their observational characteristics is provided in Table~\ref{tab:LHAASO_mq_list}.
    
% arxiv version
    \begin{table*}[ht]
        \centering
        \fontsize{7.35pt}{10pt}\selectfont
        %\scriptsize %\footnotesize   %
        \begin{tabularx}{\textwidth}{llcccccccccc}
            \hline
            \hline
            \multirow{2}{*}{LHAASO Source} & \multirow{2}{*}{Microquasar} &  \multirow{2}{*}{R.A.} &  \multirow{2}{*}{Decl.} & Extension & $\gamma$-ray Flux & \multirow{2}{*}{$\Gamma_{\rm LHA}$} & \multirow{2}{*}{$\hat{n}_s$} & \multirow{2}{*}{TS} & $\Phi^{90\%}_{0,\Gamma_\mathrm{LHA}}$ & $\Phi^{90\%}_{0,\Gamma=2.0}$ & $\Phi^{90\%}_{0,\Gamma=3.0}$ \\
             &  &  &  & (deg) & (Crab Unit) & & & & \multicolumn{3}{c}{($\rm{10^{-17} \times TeV^{-1}~cm^{-2}~s^{-1}}$)}\\
            \hline
            $\rm{J1913+0455}$ & SS 433 E. & 288.25 & 4.95 & PS & 0.10 & 2.78 & 8.78 & 0.65 & 2.77 & 7.46 & 1.28\\
            $\rm{J1910+0509}$ & SS 433 W. & 287.55 & 5.15 & PS & 0.082 & 2.92 & 11.30 & 1.09 & 1.71 & 3.31 & 1.27 \\
            $\rm{J1911+0510}$ & SS 433 central & 287.89 & 5.22 & 0.32 & 0.32 & 4.03 & 5.71 & 0.15 & 0.006 & 7.57 & 1.65 \\
            $\rm{J1819-2541}$ & V4641 Sgr & 274.82 & -25.68 & 0.36 & 3.9 & 2.67 & 0.00 & 0.00 & 477.12 & 236.15 & 538.41  \\
            $\rm{J1915+1052}$ & GRS 1915+105 & 288.63 & 10.83 & 0.33 & 0.17 & 3.07 & 0.00 & 0.00 & 1.02 & 3.64 & 1.44 \\
            $\rm{J1821+0723}$ & MAXI J1820+070 & 275.20 & 7.43 & PS & 0.13 & 3.19 & 0.00 & 0.00 & 0.30 & 2.79 & 0.74 \\
            $\rm{J1958+3522}$ & Cygnus X-1 & 299.48 & 35.37 & PS & $<$0.01 & 4.07 & 0.00 & 0.00 & 0.007 & 7.56 & 1.37 \\
            \hline
        \end{tabularx}
        \vspace{10pt}
        \begin{minipage}{\textwidth}
        \footnotesize % Optional: keeps the font size consistent with the table
        \caption{Properties of microquasars spatially associated with LHAASO gamma-ray sources and the corresponding results from a point-source neutrino search.
        Columns 1-6 list the source identifier, name, coordinates (R.A., Decl.), extension, and the $\gamma \text{-}$ray flux at $100 \mathrm{TeV}$ in Crab units (CU; $\simeq10^{-12} \mathrm{erg~cm^{-2}s^{-1}}$).
        Columns 7-9 show the results of a neutrino search where the spectral index is fixed to the value measured by LHAASO ($\Gamma_{\rm LHA}$), the estimated number of signal events ($\hat{n}_s$), and the Test Statistic (TS) value.
        The final three columns show the 90\% C.L. upper limits on the time-integrated neutrino flux normalization ($\Phi_0^{90\%}$) for different power-law spectral assumptions: fixing $\Gamma$ to the LHAASO value ($\Gamma_{\rm LHA}$), and to benchmark values of 2.0, 3.0.
        }
        \label{tab:LHAASO_mq_list}
        \end{minipage}
    \end{table*}

    \begin{enumerate}[leftmargin=*]
        \item SS433 is a microquasar with a pair of jets launched from the central BH, exhibiting distinct morphologies in different energy bands. LHAASO detected three components associated with this source. In the $25\text{-}100~\mathrm{TeV}$ range, two point-like sources (LHAASO J1913+0455 and J1910+0509) are identified, spatially coinciding with the east and west lobes; above $100~\mathrm{TeV}$, an extended source (LHAASO J1911+0513) appears near the central BH, spatially coincident with a giant atomic gas cloud ($\sim 2.5\times10^4~M_\odot$), suggesting a hadronic origin via $pp$ interactions. Leptonic models cannot reproduce the high-energy spectrum due to Klein–Nishina suppression, further strengthening the hadronic interpretation.

        \item V4641 Sgr is associated with an extended gamma ray source (LHAASO J1819–2541), with photon energies reaching $\sim800~\mathrm{TeV}$. The source exhibits no spectral cutoff and no alternative astrophysical counterpart is identified nearby. %The spectral index of $2.84 \pm 0.17$ matches previous measurements by HAWC. 
        The observed features point to efficient hadronic acceleration, possibly extending beyond 10 PeV, while leptonic models are disfavored due to Klein–Nishina suppression. 

        \item GRS 1915+105 is associated with an extended source (LHAASO J1914+1049), which shows emission up to $1~\mathrm{PeV}$. The gamma-ray morphology aligns with the axis connecting the radio lobes observed by ALMA at $\sim$50 pc from the central source. Although the angular resolution limits spatial identification at its distance of 9.4 kpc, the emission could originate from jet–medium interactions, consistent with hadronic models.

        \item MAXI J1820+070 is associated with a point-like UHE source (LHAASO J1821+0726), with gamma rays observed up to 400 TeV. The source is offset by $0.27^\circ$ from the black hole and aligns with the axis of the receding jet ejected during its 2018 outburst. The positional and temporal correlation suggests that the emission may be linked to episodic jet activity. Intriguingly, the error region of the IceCube high-energy neutrino alert IC250813A encompasses the position of MAXI J1820+070. This spatial coincidence provides strong additional motivation to investigate this source as a potential site of hadronic acceleration and neutrino production.

        \item Cygnus X-1 is associated with the point-like source (LHAASO J1957+3517), showing gamma ray emission extending to $\sim100$ TeV. The source is offset by $0.19^\circ$ from the binary and aligns with the arc-shaped radio bow structure inflated by the jet. Although the spectrum is relatively soft, the morphology supports a jet-driven origin, consistent with a hadronic interpretation.
    \end{enumerate}
    
    \section{Analysis Method}\label{sec:analyse}
    To search for a neutrino signal from the target source, we perform an unbinned maximum likelihood analysis following the methodology described in \citet{Braun_method_2008,IceCube_method_2011}. Assuming a power-law neutrino spectrum $dN_\nu/dE_\nu \propto E_\nu^{-\Gamma}$, the likelihood function is defined as
    \begin{flalign}
        & \mathcal{L}(n_s)=\prod_k \prod_{i}^{N_k} \left [ \frac{n_s^k}{N_k}S_i+ 
        \left( 1-\frac{n_s^k}{N_k} \right)B_i \right ], &
        \label{eq:like}
    \end{flalign}
    where $S_i$ and $B_i$ are the signal and background probability density functions (PDFs) for event $i$, $N_k$ and $n_s^k$ denote the total and signal event counts for sample $k$, respectively. The per-sample signal count is parameterized as a fraction of the total signal: 
    \begin{flalign}
        & n_s^k = f_k \times n_s. &
    \end{flalign} 
    where $f_k$ represents the expected fractional contribution of sample $k$ to the total signal. 
    
    For a single-source analysis, $f_k$ is determined by detector acceptance $\omega_{j, \rm acc}$.
    \begin{flalign}
        & f_k = \frac{\omega_{j, \rm acc}^k}{\sum_{k^\prime} \omega_{j,\rm acc}^{k^\prime}} = \frac{t_k \int E_\nu^{-\Gamma} A_{\rm eff}^k(E_\nu,\delta_j) dE_\nu}{\sum_{k^\prime} t_{k^\prime} \int E_\nu^{-\Gamma} A_{\rm eff}^{k^\prime}(E_\nu,\delta_j) dE_\nu} &
    \end{flalign}
    Here, $t_k$ is the livetime of sample $k$, and $A^k_{\rm eff}(E_\nu, \delta_j)$ is its effective area, which is a function of neutrino energy $E_\nu$ and source declination $\delta_j$. 

    For a stacked (multi-source) analysis, the calculation is extended by introducing an additional source-specific weight $\omega_{j,\rm model}$, which represents the hypothesized relative contributions of each source $j$.
    \begin{flalign}
        & f_k = \frac{\sum_j \omega_{j,\rm model}~t_k \int E_\nu^{-\Gamma} A_{\rm eff}^k(E_\nu,\delta_j)dE_\nu}{\sum_{k^\prime} \sum_j  \omega_{j, \rm model}~t_{k^\prime} \int E_\nu^{-\Gamma} A_{\rm eff}^{k^\prime}(E_\nu,\delta_j)dE_\nu} &
    \end{flalign} %= \frac{\sum_j \omega_{j,\rm m} \omega_{j,\rm acc}^k}{\sum_{k^\prime}\sum_j \omega_{j,\rm m} \omega_{j,\rm acc}^{k^\prime}} 

    The signal and background PDFs are constructed by independent spatial and energy terms for each event $i$:
    \begin{flalign}
        & S_i = S^{\mathrm{spat}}(\bm{x}_i,\sigma_i,\bm{x}_j,\sigma_j) S^{\mathrm{ener}}(E_i,\bm{x}_j,\Gamma) &
    \end{flalign}
    \begin{flalign}
        & B_i = B^{\mathrm{spat}}(\delta_i) B^{\mathrm{ener}}(E_{\mathrm{rec}}, \delta_i) &
    \end{flalign}
    
    The spatial signal term describes the distribution of the reconstructed event directions $\bm{x}_i$, and is modeled as a 2D Gaussian centered on the source position $\bm{x}_j$:
    \begin{flalign}
        & S^{\mathrm{spat}}(\bm{x}_i,\sigma_i,\bm{x}_j,\sigma_j)=\frac{1}{2\pi (\sigma_i^2+\sigma_j^2)}\exp\left[ -\frac{D(\bm{x}_i,\bm{x}_j)^2}{2(\sigma_i^2+\sigma_j^2)}\right], & 
        \label{eq:s_spat}
    \end{flalign}
    where $D\left( \bm{x}_i, \bm{x}_j \right)$ is the angular separation between the event and the source, $\sigma_i$ is the event angular uncertainty, and $\sigma_j$ represents the source extension.

    The energy signal term describes the probability of reconstructing an energy $E_i$ given a true neutrino energy $E_\nu$, and is constructed from the detector effective area $A_{\rm eff}^k$ and the energy response function $\mathcal{M}_k$:
    \begin{flalign}
        & S^{\mathrm{ener}}(E_i,\bm{x}_j,\Gamma)=\frac{\int E_\nu^{-\Gamma}A_{\rm{eff}}(E_\nu,\delta_j)\mathcal{M}_k(E_i|E_\nu,\delta_j)dE_\nu }{\int E_\nu^{-\Gamma}A_{\rm{eff}}(E_\nu,\delta_j)dE_\nu}, &
        \label{eq:s_ener}
    \end{flalign}
    where $\delta_j$ is the declination of the source and $\Gamma$ is the assumed spectral index.

    The background PDF depends primarily on the event declination, assuming uniformity in right ascension. The spatial background term is estimated by counting events in a declination band:
    \begin{flalign}
        & B^{\mathrm{spat}}(\delta_i) = \frac{\sum_j N_{ij}^k}{N_k  \Omega_{\delta_i \pm 3^\circ}}, & 
        \label{eq:b_spat} 
    \end{flalign}
    where $\Omega_{\delta_i \pm 3^\circ}$ is the solid angle of a $6^\circ$ declination band centered at $\delta_i$, $N_{ij}^k$ is the number of events in that band, and $N_k$ is the total number of events in sample $k$. The energy background term is constructed using small bins in declination and reconstructed energy:
    \begin{flalign}
        & B^{\mathrm{ener}}(E_{\mathrm{rec}}, \delta_i) = \frac{N_{ij}^k}{\sum_j N_{ij}^k}, & \label{eq:b_ener}
    \end{flalign}
    where the bin is defined by $\sin \delta \in [\sin \delta_i, \sin \delta_i + 0.02]$ and $\log_{10} E_{\mathrm{rec}} \in [\log_{10} E_j, \log_{10} E_j + 0.1)$.
    
    To evaluate the presence of a source at a given position, we define the test statistic (TS) as twice the logarithm of the likelihood ratio between the best-fit signal hypothesis and the null (background-only) hypothesis:
    \begin{flalign}
        & \mathrm{TS} = 2 \left[ \log \mathcal{L}(\hat{n}_s) - \log \mathcal{L}(n_s \text{=} 0) \right]. & 
        \label{eq:ts}
    \end{flalign}
    Given a spectral index $\Gamma$, we compute $\hat{n}_s$ and the corresponding TS. If $\rm{TS<9}$, no significant signal is considered to be present. In this case, the 90\% confidence level upper limit on the signal count, $n_{90}$, is derived using the likelihood fitting procedure. The corresponding 90\% upper limit on the neutrino flux, $\Phi_{\nu_\mu+\bar{\nu}_\mu}^{90\%}$, is obtained via
    \begin{flalign}
        & n_{\rm 90} = \sum_k \int dt_k \int \Phi_{\nu_\mu+\bar{\nu}_\mu}^{90\%} A_{\rm eff}^k(E_\nu,\delta_j) dE_\nu. &
    \end{flalign}
    
    \section{Results and Discussion}\label{sec:results}
    This section presents the results of our analysis of 10-year IceCube data, following the methodology described in Section~\ref{sec:analyse}. We first report the results of the single source search in Section~\ref{subsec:single_result} and use the upper limits derived to constrain the hadronic gamma ray flux. Subsequently, in Section~\ref{subsec:stack_result}, we performed a stacking analysis to evaluate the collective contribution of all targeted microquasars to the diffuse neutrino flux along the Galactic Plane.
    
    \subsection{Single Source Analysis and Hadronic Constraints}\label{subsec:single_result}
    \begin{figure*}
        \centering
        \includegraphics[width=0.24\textwidth]{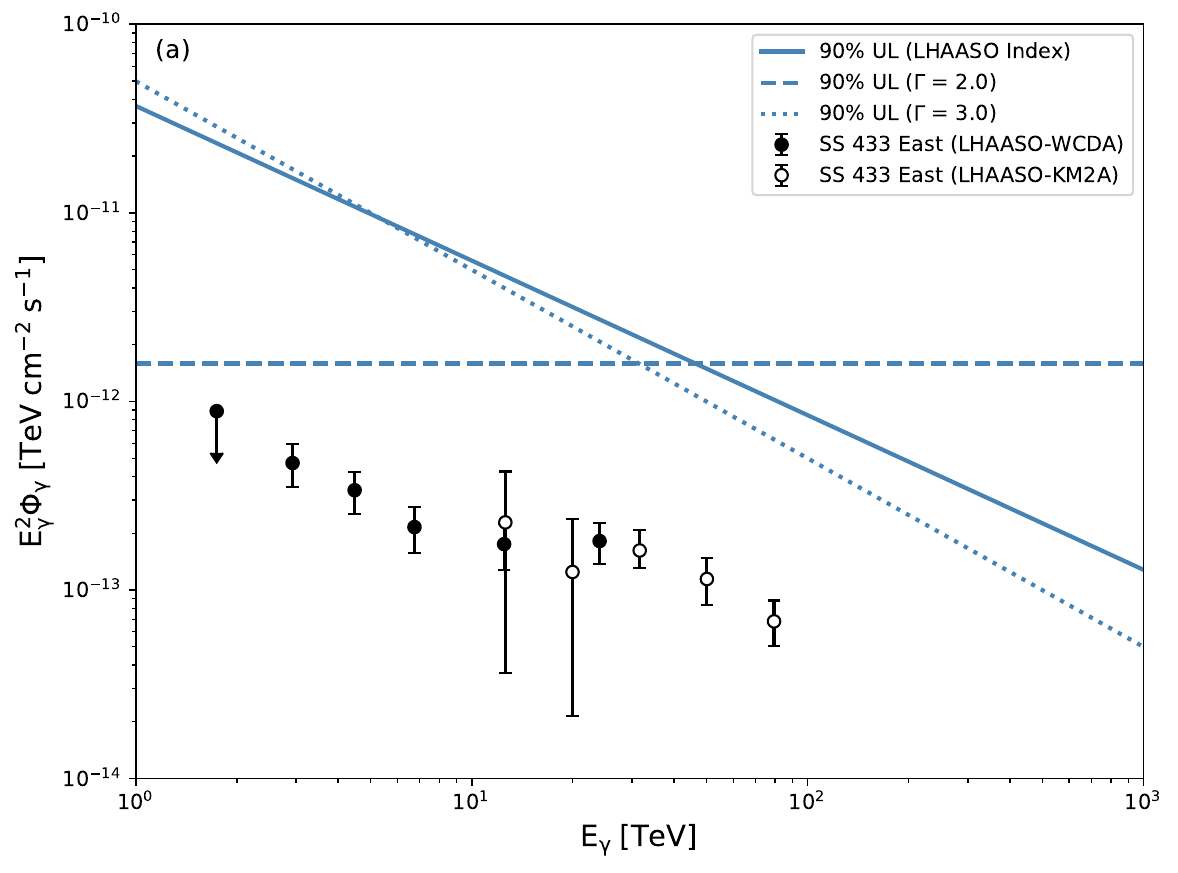}
        \includegraphics[width=0.24\textwidth]{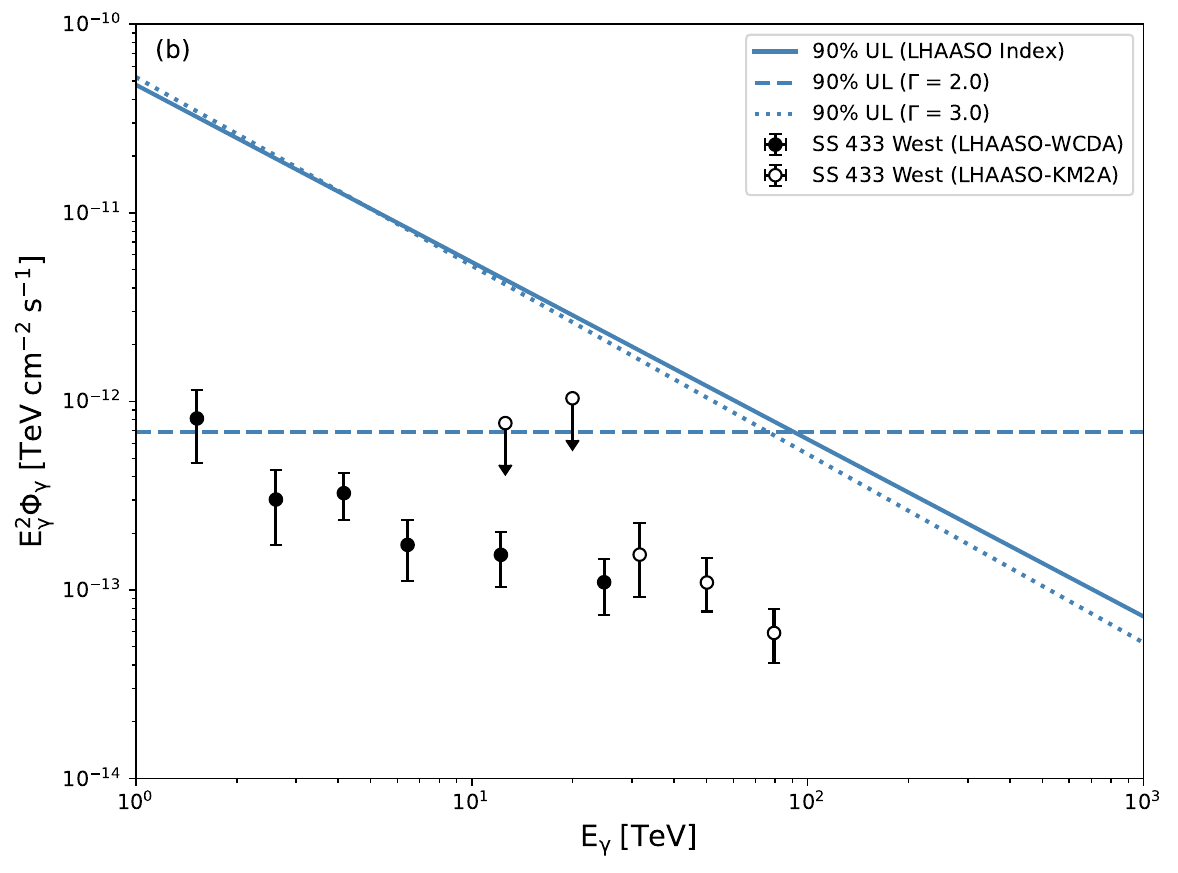}
        \includegraphics[width=0.24\textwidth]{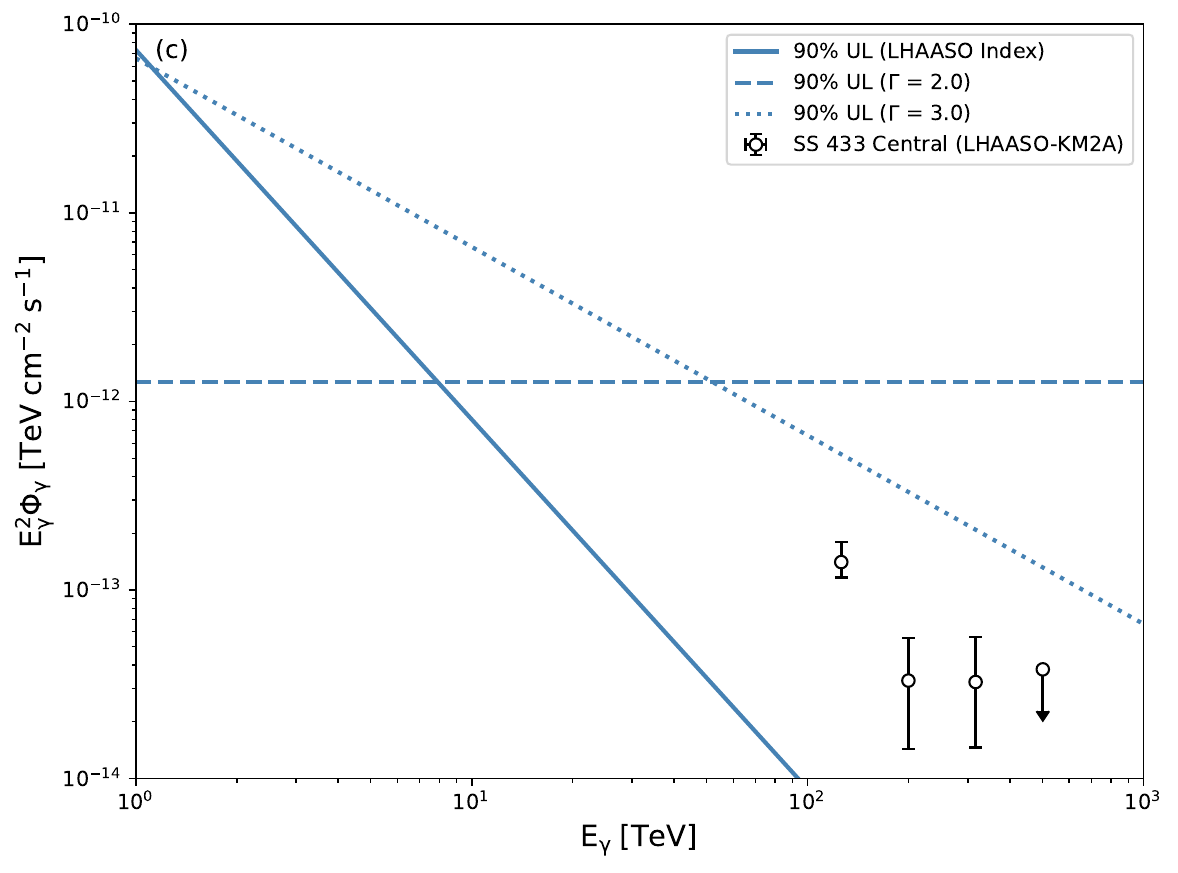}
        \includegraphics[width=0.24\textwidth]{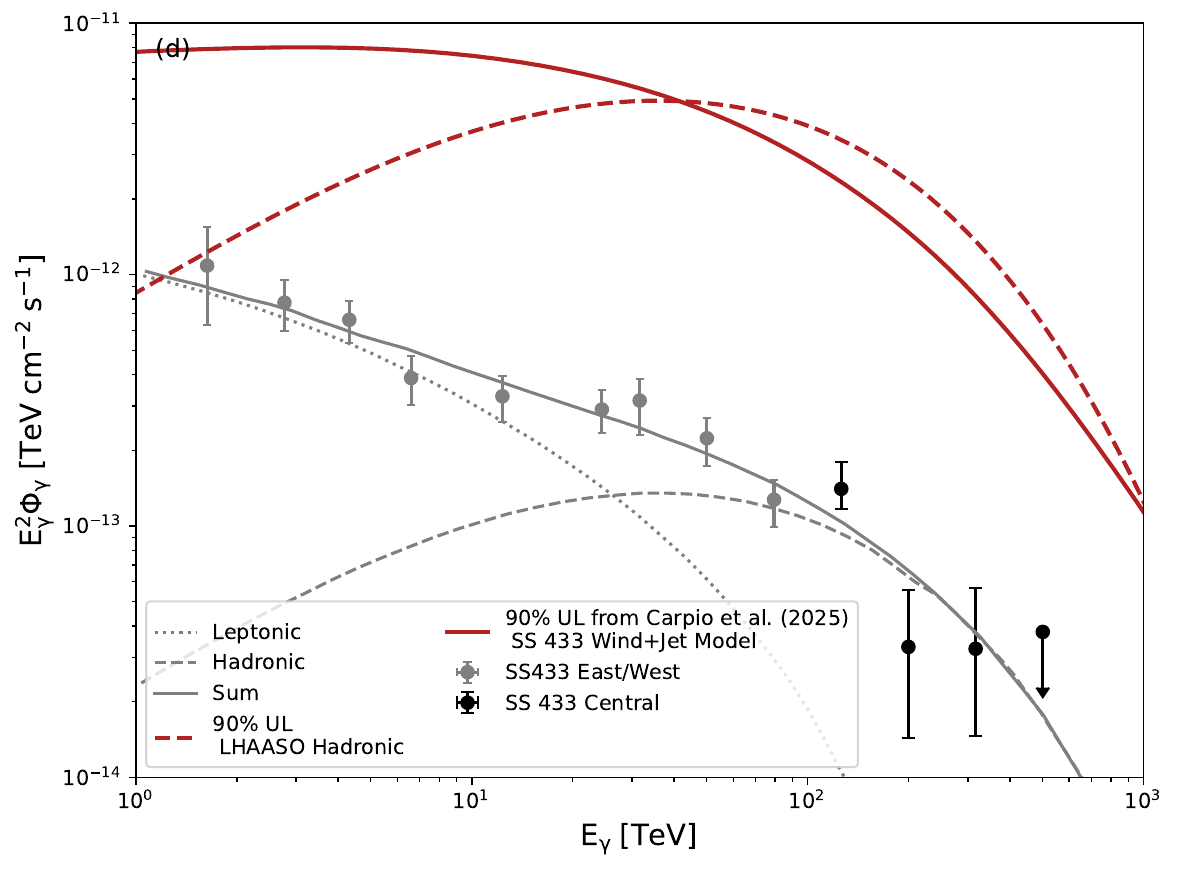}
        \includegraphics[width=0.24\textwidth]{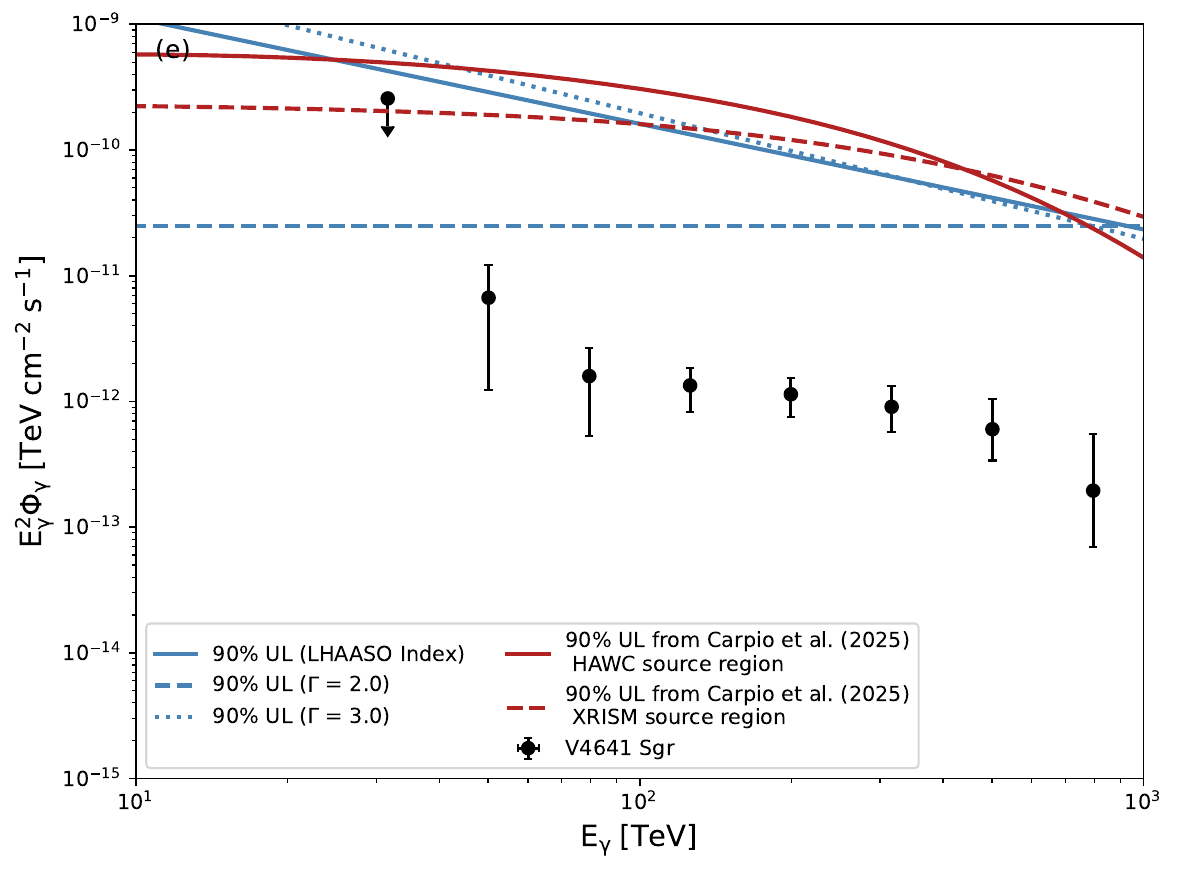}
        \includegraphics[width=0.24\textwidth]{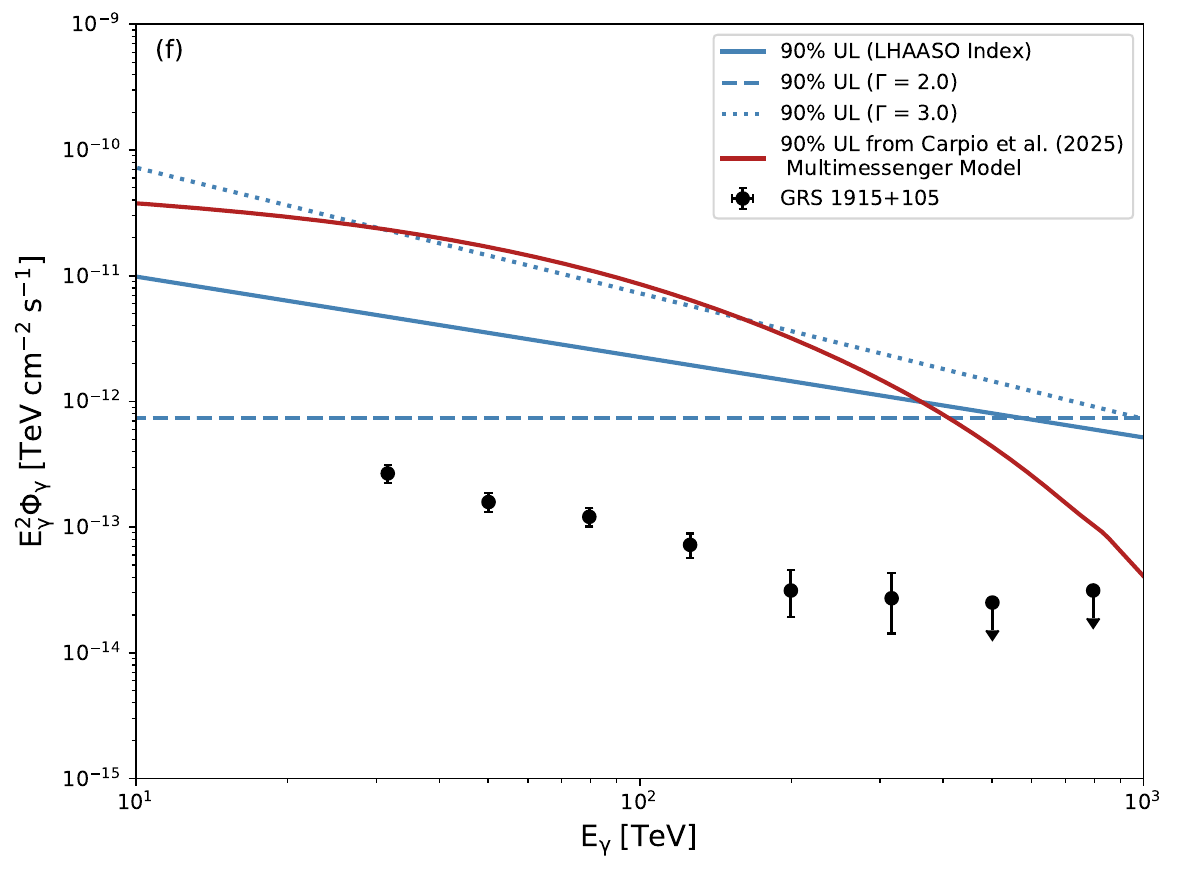}
        \includegraphics[width=0.24\textwidth]{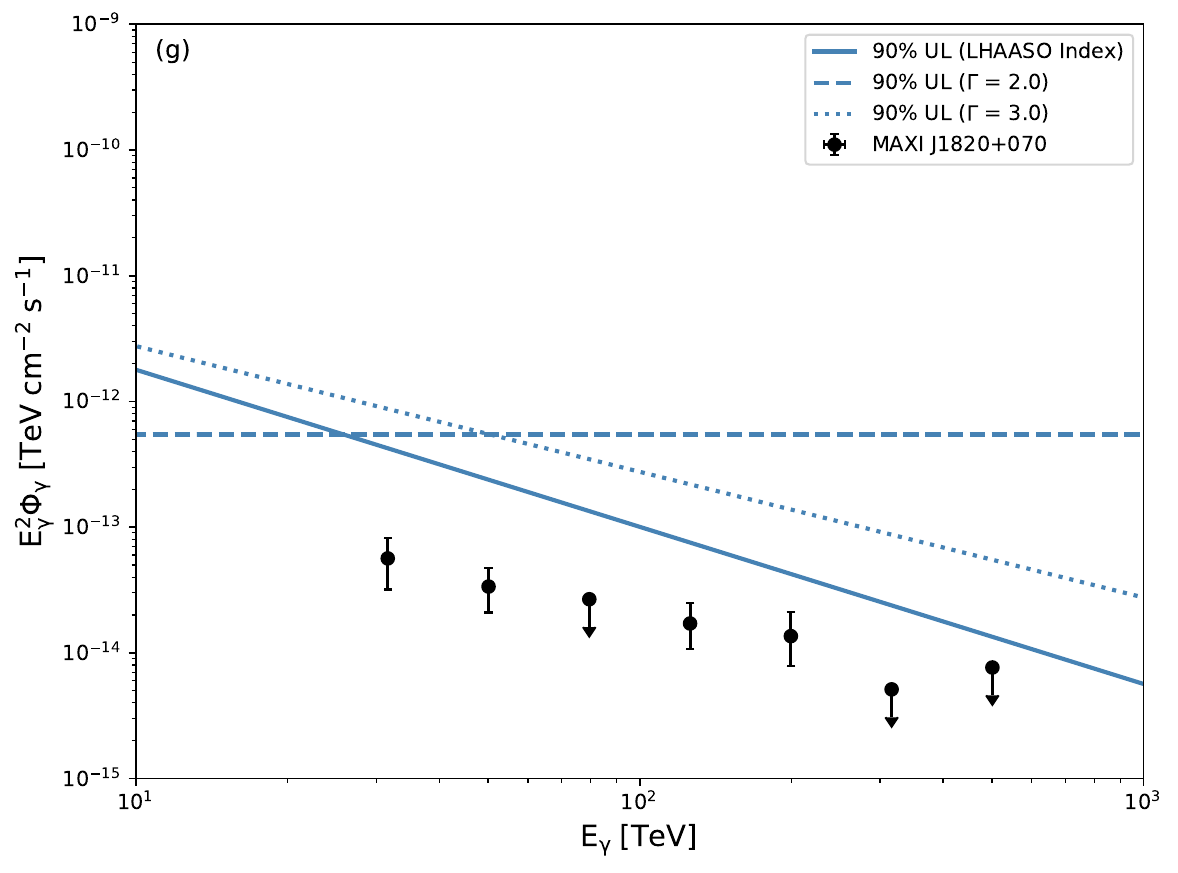}
        \includegraphics[width=0.24\textwidth]{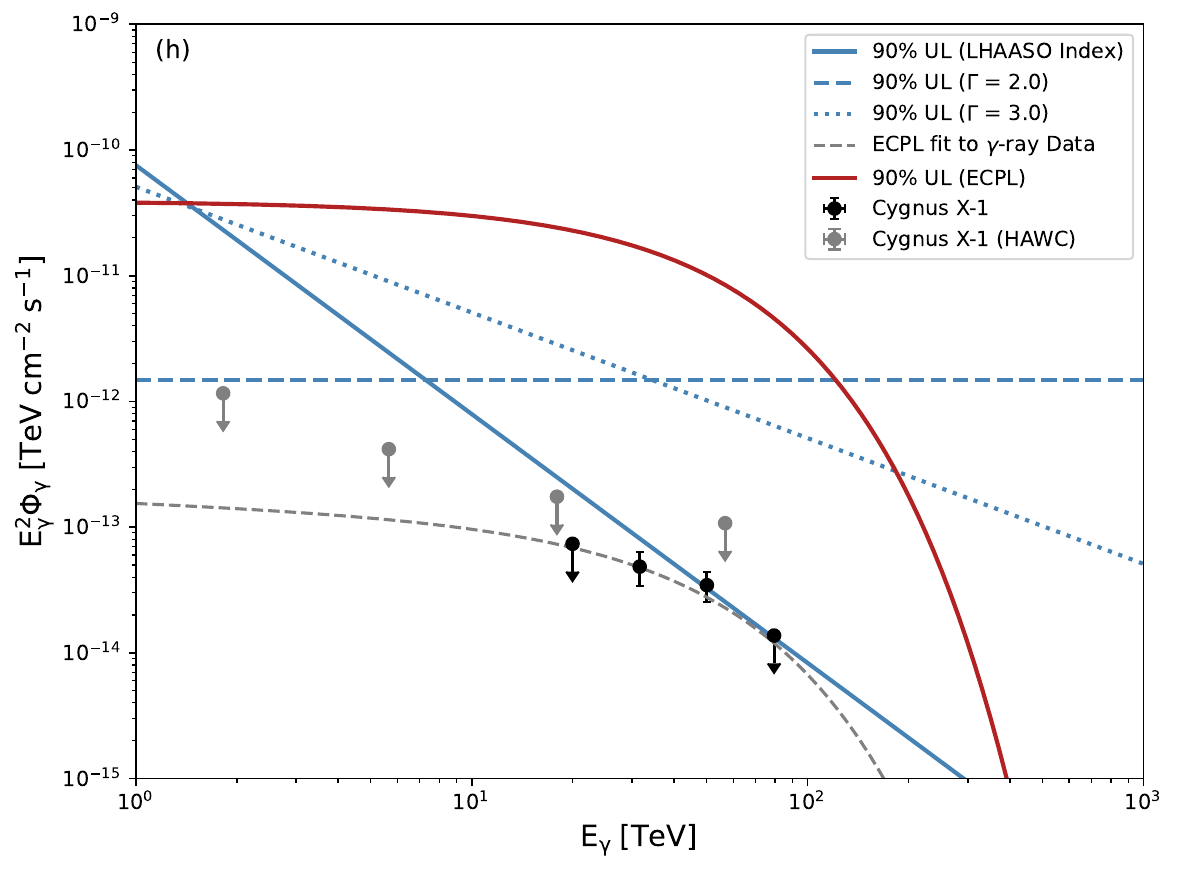}
        \caption{Constraints on the hadronic gamma-ray flux for individual microquasars. In each panel, the black points show the gamma-ray flux measured by LHAASO, with 1$\sigma$ statistical error bars. 
        The blue lines represent the $90\%$ C.L. upper limits on the hadronic flux, derived assuming a power-law spectrum with the LHAASO best-fit index (solid), and benchmark indices of $\Gamma=2.0$ (dashed) and $\Gamma=3.0$ (dotted). 
        The red curves represent constraints from source-specific, physically motivated models. 
        Specifically, panel (d) compares two models for SS 433, one based on the LHAASO molecular cloud interaction scenario (red dashed line) and the other based on the "wind+jet" model of \citep{Carpio:2025arz} (red solid line). 
        For V4641 Sgr in panel (e), the red dashed and solid lines correspond to the lepto-hadronic (XRISM region) and purely hadronic (HAWC region) scenarios from \citep{Carpio:2025arz}, respectively. 
        The limit for GRS 1915+105 in panel (f) is derived from the purely hadronic model from the same work. 
        Finally, in panel (h) for Cygnus X-1, the red curve is the limit derived using an ECPL spectral shape, with the ECPL model fit itself shown as the gray dashed line.}
        \label{fig:hadronic}
    \end{figure*}

    The recent detection of UHE gamma-ray emission from five galactic microquasars by LHAASO, with spectra characterized by power-law indices $\Gamma_{\rm LHA}$, provides a strong motivation for a targeted neutrino search. We tested the hypothesis of a hadronic origin for this emission, which would imply a correlated neutrino flux with a similar power-law spectrum.
    For each microquasar, we performed an unbinned likelihood analysis assuming distinct spectral hypotheses. First, we fixed the neutrino spectral index $\Gamma$ to the best-fit value measured by LHAASO ($\Gamma_{\rm LHA}$). In addition, we tested two benchmark indices of $\Gamma=2.0$ and $\Gamma=3.0$. No statistically significant excess of neutrinos is found from any individual source for any of the spectral assumptions tested. The best-fit number of signal events ($\hat{n}_s$) and the corresponding TS values for the analysis using the best-fit indices derived by LHAASO are summarized in Table~\ref{tab:LHAASO_mq_list}. Given the non-detection, we calculated the $90\%$ confidence level (C.L.) upper limits on the time-integrated neutrino flux normalization, $\Phi^{90\%}_{0,\Gamma}$, at $100~\mathrm{TeV}$ for each spectral assumption.
These limits constrain the hadronic contribution to the LHAASO-observed gamma-ray emission, based on the direct physical connection between neutrino and gamma-ray fluxes from pion decay in hadronic interactions:
    \begin{flalign}
        & \frac{1}{3}\frac{dN_\nu}{dE_\nu}= K_\pi \frac{dN_\gamma}{dE_\gamma}, &
    \end{flalign}
    where the energy relation is approximately $E_\gamma = 2E_\nu$, and $K_\pi$ is the ratio of charged to neutral pions produced. This ratio depends on the specific hadronic interaction channel: $K_\pi \simeq 2$ for proton-proton ($pp$) collisions and $K_\pi \simeq 1$ for photohadronic ($p\gamma$) interactions. In this work, we focus on the $pp$ collision scenario. Using this relation, we translate the $90\%$ C.L. upper limits on the neutrino flux into the corresponding upper limits on the gamma-ray flux that could originate from the hadronic channel. These constraints are compared with the total observed gamma-ray emission in panels (a)-(c) and (e)-(h) of Figure~\ref{fig:hadronic}. The limit derived using the LHAASO best-fit index, and the benchmark indices $\Gamma=2.0$ and $\Gamma=3.0$, are shown as blue solid, dashed and dotted lines, respectively.

    The simple power-law treatment is inadequate for sources whose LHAASO spectra are poorly constrained or exhibit complex features. Accordingly, we employ source-specific, physically motivated spectral models for several sources to enable a more direct comparison with theoretical predictions. 
    For SS 433, we tested two hadronic scenarios. First, we adopted the spectral shape from the LHAASO Collaboration's model, which attributes the UHE gamma-ray emission to cosmic-ray interactions with a molecular cloud, as a template for the neutrino spectrum. Second, we tested the composite "wind+jet" model from \citep{Carpio:2025arz}, which combines a lepto-hadronic jet with a hadronic wind component. The resulting 90$\%$ C.L. upper limits on the hadronic gamma-ray flux for these two models are shown as red dashed and solid lines, respectively, in panel (d) of Figure~\ref{fig:hadronic}.
    For Cygnus X-1, given that its LHAASO spectrum is constrained by only two data points, we incorporate lower energy upper limits from HAWC \citep{Ohira:2024qtr} and refit the combined data set with an Exponential Cutoff Power Law (ECPL) model (shown as the gray dashed line in panel (h) of Figure~\ref{fig:hadronic}). We then used this ECPL spectral shape as the neutrino spectrum in our analysis to derive the $90\%$ C.L. upper limits on the neutrino flux, which is subsequently converted to a limit on the hadronic gamma ray flux, shown as the red curve in the same panel. 
    Finally, we extend the analysis to the extended sources V4641 Sgr and GRS 1915+105 by applying the source-specific models from \citep{Carpio:2025arz}. For V4641 Sgr, we tested two proposed scenarios: a lepto-hadronic model for the compact (10 pc) XRISM source region and a purely hadronic model for the extended (60 pc) HAWC source region. The resulting upper limits are shown as red dashed and solid lines, respectively, in panel (e). For GRS 1915+105, we adopt the purely hadronic model from the same work, with the resulting upper limit shown as the red curve in panel (f).
    
    \subsection{Stacking Analysis}\label{subsec:stack_result}
    \begin{figure}
        \centering
        \includegraphics[width=1\linewidth]{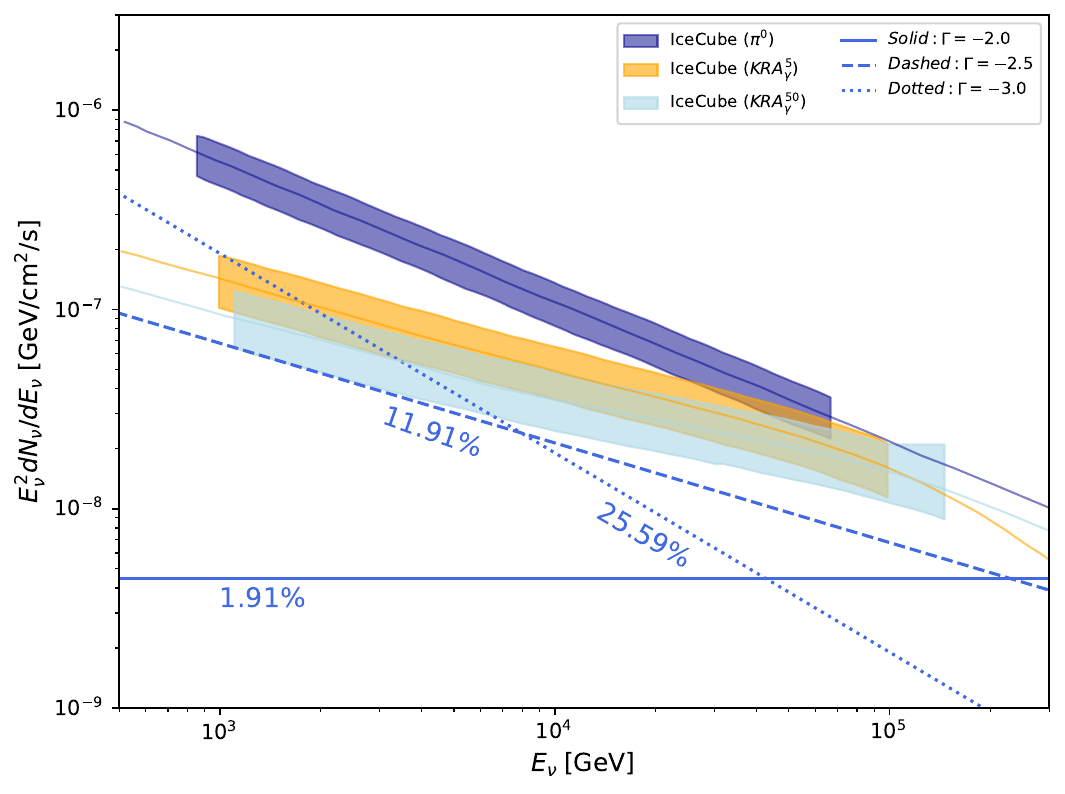}
        \caption{The 90$\%$ C.L. upper limits on the stacked, all-flavor neutrino flux from the galactic microquasar populations. The limits are shown for three different power-law spectral index assumptions: $\Gamma=2.0$ (solid line), $\Gamma=2.5$ (dashed line), and $\Gamma=3.0$ (dotted line). The constraints are compared to the galactic plane neutrino flux for each Galactic plane models, which are the $\pi^0$ (purple), $\rm{KRA}_\gamma^5$ (orange) and $\rm{KRA}_\gamma^{50}$ (light blue) while solid lines and shaded regions indicates the best-fitting and $1\sigma$ uncertainties. The number above each line displays the fraction can contribute to the galactic plane neutrino flux fitted by $\pi^0$ template.}
        \label{fig:stack_mq}
    \end{figure}
    
    To assess the cumulative contribution of the known microquasar population to the Galactic Plane neutrino flux, we perform a stacking analysis incorporating a census of all identified microquasars. We assume that all sources follow a similar power-law energy spectrum (i.e., the same spectral index) and contribute equally to the total neutrino emission (i.e., equal weighting). 
    For this analysis, we convert the $\nu_\mu + \bar{\nu}_\mu$ flux limits derived from the muon track data to all-flavor neutrino fluxes by a factor of 3. We then test three benchmark power-law hypotheses with spectral indices of $\Gamma =$ 2.0, 2.5, and 3.0. The resulting $90\%$ C.L. upper limits on the stacked all-flavor neutrino are presented in Figure~\ref{fig:stack_mq}. 
    These constraints are compared with existing Galactic Plane diffuse neutrino models, including predictions $\pi^0$ decay (dark blue), $\mathrm{KRA}_\gamma^5$ (orange), and $\mathrm{KRA}_\gamma^{50}$ (light blue) \citep{IceCube_GP_diffuse}. To quantify the constraining power of our analysis, we evaluate our limits relative to the benchmark $\pi^0$ decay model, our upper limits correspond to a matrix indicating that the microquasar population contributes no more than 1.91$\%$, 11.91$\%$, and 25.59$\%$ for $\Gamma=2.0, 2.5 $ and $3.0$, respectively. 
    %Therefore, our analysis limits the potential contribution of the microquasar population to the Galactic neutrino flux originating from $\pi^0$ decay, suggesting that other sources are likely required to explain the total observed flux.
    Therefore, our analysis demonstrates that microquasars can account for only a fraction of the Galactic neutrino flux attributable to decay of $\pi^0$, which implies that a significant portion must therefore originate from other source populations.
    
    \section{Conclusion and Outlook}\label{sec:outlook}
    \begin{figure*}
        \centering
        \includegraphics[width=0.32\linewidth]{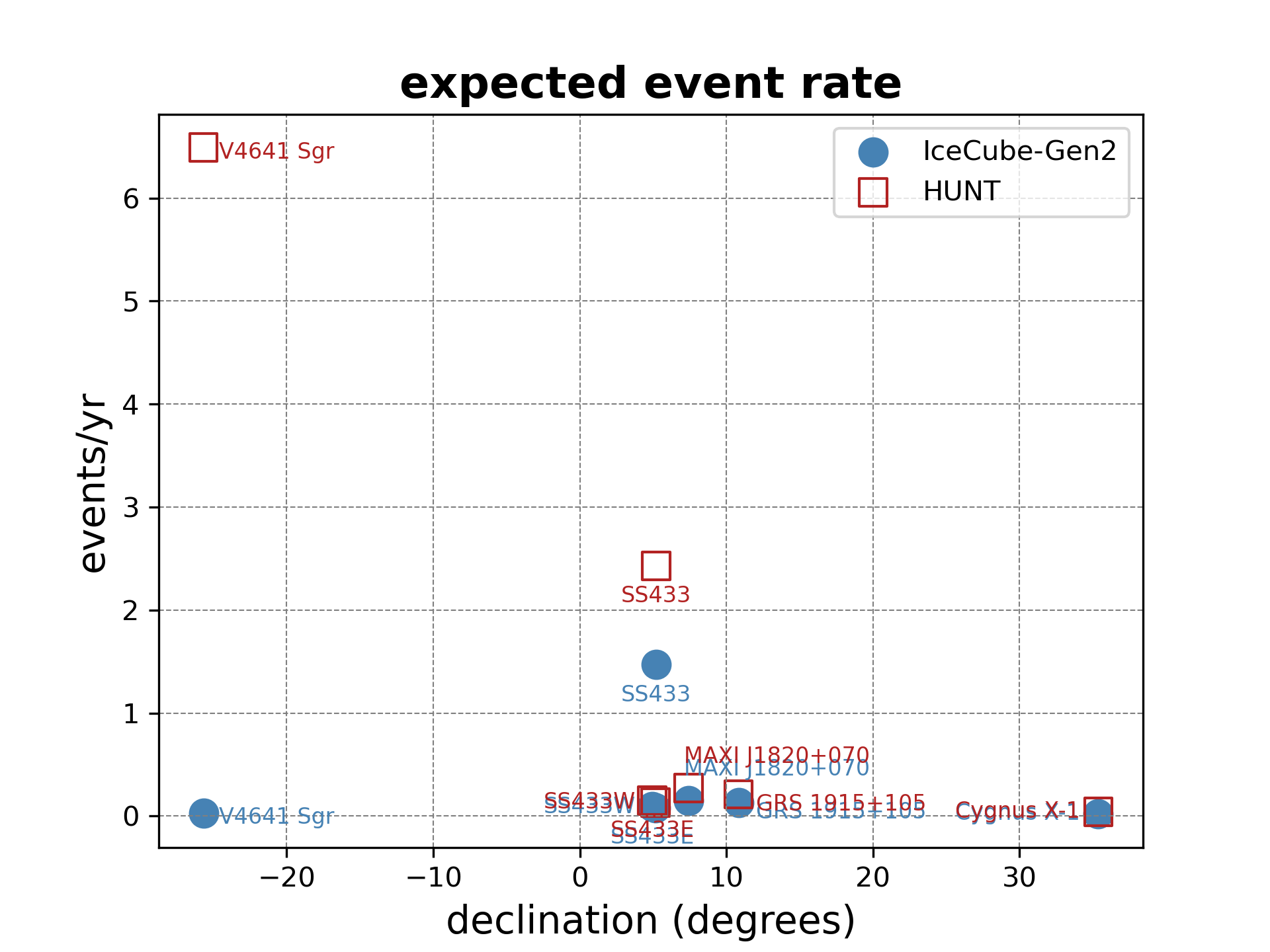}
        \includegraphics[width=0.32\linewidth]{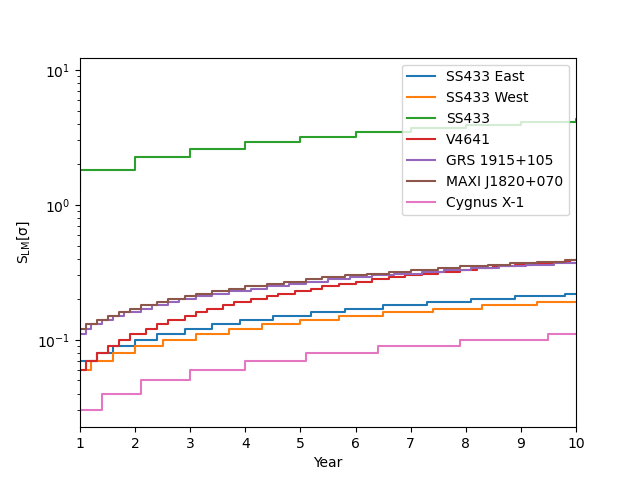}
        \includegraphics[width=0.32\linewidth]{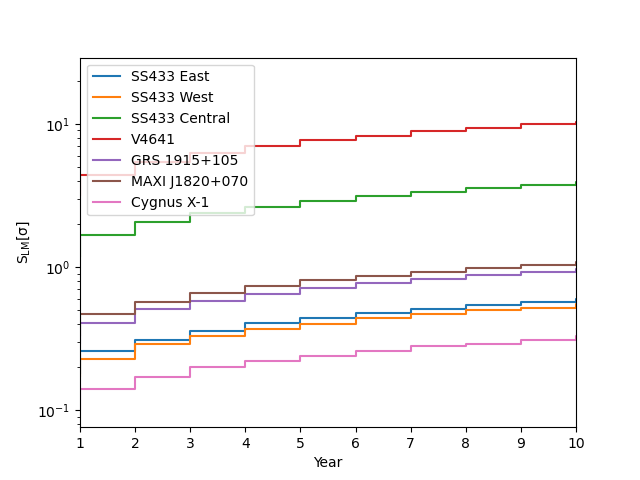}
        \caption{Left: Expected neutrino event rates from microquasars, calculated for IceCube-Gen2 (blue solid circles) and HUNT (red open squares). The rates are based on the neutrino flux converted from the LHAASO gamma-ray measurements. Middle: Cumulative statistical significance $\rm{S_{LM}~(\sigma)}$ for a neutrino signal ($E_\nu > 10 \rm TeV$) as a function of observation time (in years) for IceCube-Gen2. Right: Same as the middle panel, but for the HUNT detector.}
        \label{fig:Gen2_ns}
    \end{figure*}
    
    In this work, we conducted a comprehensive search for high-energy neutrino emission from LHAASO-observed microquasars using ten years of IceCube data. Our analysis, which included single-source and population stacking approaches, did not find a statistically significant signal. We used the resulting upper limits on the neutrino flux to place source-specific constraints on the hadronic gamma-ray emission. This non-detection is consistent with the current sensitivity limitations of neutrino astronomy for Galactic sources.

    The detection capabilities for high-energy neutrinos are expected to improve significantly with the next generation of observatories. At the South Pole, IceCube-Gen2 will provide an effective area approximately 7.5 times larger than that of the current IceCube detector \citep{IceCube-Gen2:2020qha, Schumacher:2021hhm}. We predicted the future detectability of these microquasars with next-generation observatories, specifically IceCube-Gen2 \citep{IceCube-Gen2_2020} and the High-energy Underwater Neutrino Telescope (HUNT) \citep{HUNT:2023mzt}. These predictions are based on the neutrino fluxes derived under the assumption that the entire LHAASO gamma-ray flux is of hadronic origin. We calculate the expected number of signal events by convolving these fluxes with the respective detector responses. Figure~\ref{fig:Gen2_ns} summarizes these predictions, with the left panel showing the event rates of neutrinos from the five microquasars at different declinations for both detectors, and the middle and right panels illustrating the evolution of detection significance over the operational time for IceCube-Gen2 and HUNT, respectively, at a 10 TeV energy threshold.
    The significance of the detection is calculated using the Li–Ma formula \citep{li1983analysis}, considering the background of atmospheric and diffuse astrophysical neutrinos \citep{IceCube:2020acn}. The background level is estimated within a region defined by the point spread function (PSF) with $\sigma_{\mathrm{PSF}} = 0.3^\circ$ \citep{Bradascio:2019gyw,HUNT_2025icrc}.

Our predictions highlight the complementary roles of next-generation detectors. For IceCube-Gen2, detecting most of the LHAASO-observed  microquasars may require substantial exposure. However, HUNT's northern location could allow it to detect V4641 Sgr within a year, if the UHE gamma rays from the source are purely hadronic. 
Conversely, the non-detection of neutrinos from V4641 Sgr by HUNT would challenge the assumption of a purely hadronic origin for the UHE gamma rays, thereby placing stringent constraints on the hadronic contribution and refining our understanding of the underlying particle acceleration mechanisms.
Together, they will provide a powerful test of hadronic emission from these systems.

    \section*{Acknowledgments}
    We thank Tianqi Huang for the useful comments.
    H.N. He is supported by Project for Young Scientists in Basic Research of Chinese Academy of Sciences (No. YSBR-061), and by NSFC under the grants No.12173091, No.12333006, and No.12321003, and the Strategic Priority Research Program of the Chinese Academy of Sciences No.XDB0550400.

\end{CJK*}
    \bibliography{MQ_refer.bib}{}
    \bibliographystyle{aasjournal}
\end{document}